\documentstyle[10pt,twoside]{siamltex}
\newfont{\bb}{msbm10}
\def\Bbb#1{\mbox{\bb #1}}
\def\reals{ {\Bbb R}}
\def\complex{ {\Bbb C}}

\def\quaternions{ {\Bbb H}}

\def\nxn{\mbox{\tiny{$n\times n$}}}
\def\nxm{\mbox{\tiny{$n\times m$}}}
\def\mxn{\mbox{\tiny{$m\times n$}}}
\def\2nx2n{\mbox{\tiny{$2n\times 2n$}}}
\def\kxk{\mbox{\tiny{$k\times k$}}}
\def\nkxnk{\mbox{\tiny{$(n-k)\times (n-k)$}}}
\def\n1xn1{\mbox{\tiny{$(n-1)\times (n-1)$}}}
\def\nx1{\mbox{\tiny{$n \times 1$}}}

\def\1{\mbox{\tiny{$1$}}}
\def\2{\mbox{\tiny{$2$}}}
\def\4{\mbox{\tiny{$4$}}}
\def\-{\mbox{\tiny{$-$}}}
\def\+{\mbox{\tiny{$+$}}}
\def\S{\mbox{\tiny $S$}}

\def\cF{{\cal F}}
\def\cD{{\cal D}}

\def\inv{{\mbox{\tiny $-1$}}}

\begin{document}

\bibliographystyle{plain}
\title{
The Quaternionic Determinant\thanks{
{\em Electronic Journal of Linear Algebra} {\bf 7}, 100-111 (2000)}}

\author{
Nir Cohen\thanks{
Department of Applied Mathematics, IMECC, University of Campinas, 
CP 6065, 13081-970  Campinas (SP) Brazil
(nir@ime.unicamp.br, deleo@ime.unicamp.br). 
The authors were partially supported by CNPq grant 300019/96-3 (Nir Cohen)
and FAPESP grant 99/09008-5 (Stefano De Leo).}
\and
Stefano De Leo\footnotemark[2]
}

%

\markboth{N.\ Cohen, S.\ De Leo}
{The Quaternionic Determinant}

\maketitle

\begin{abstract}
The determinant for complex matrices cannot 
be extended to quaternionic matrices. Instead, the
Study determinant and the closely related $q$-determinant 
are widely used. We show that the Study determinant can be
characterized as the unique functional which extends
the {\em absolute value} of the complex determinant  
and discuss its spectral and linear
algebraic aspects.
\end{abstract}

\begin{keywords}
Quaternions, Matrices, Determinant.
\end{keywords}
\begin{AMS}
15A09, 15A33 \end{AMS}


\section{Introduction}

Quaternionic linear algebra is attracting growing interest in  
theoretical physics~\cite{FIN1}-\cite{BOOK2}, 
mainly in the context 
of quantum mechanics and field theory~\cite{DERO}. Quaternionic mathematical 
structures have recently appeared in discussing eigenvalue 
equations~\cite{EIN1,EIN2}, group theory~\cite{GT1,GT2} and grand 
unification model~\cite{UM1,UM2} within a quaternionic formulation 
of quantum physics. 

The question of extending the determinant from
complex to quaternionic matrices has been considered
in the physical literature~\cite{BOOK1}-\cite{DERO}. 
The possibility of such an extension
has been contemplated by Cayley~\cite{CAY},
without much success, as early as 1845. A canonical
determinant functional was introduced by Study~\cite{STU}
and its properties axiomatized by Dieudonn\'e~\cite{DIE}. 
The details can be found in the excellent
survey paper of Aslasken~\cite{ASL}. Study's determinant
is denoted as $\mbox{Sdet}$, and up to a trivial 
power factor, is identical to the $q$-determinant, 
$\mbox{det}_q$, found in most of the literature~\cite{ZHA} 
and to Dieudonn\'e's determinant, denoted as Ddet. 
Study's determinant is closely related
to the q-determinant and to Dieudonne's determinant. Specifically, 
$\mbox{det}_q=\mbox{Sdet}^{\2}=\mbox{Ddet}^{\4}$.

In these works, $\mbox{Sdet}$ was considered as a {\it generalization}
of the determinant, $\mbox{det}$, in the sense that the two functionals 
share a common set of axioms. Specifically, 
$\mbox{Sdet}$ is the unique, up to a trivial power factor,
functional ${\cF}:\quaternions^{\nxn}$ which satisfies
the following three axioms:

1.~~~~~~~~ $\cF(A)=0$ if and only if $A$ is singular;

2.~~~~~~~~ multiplicativity: $\cF(AB)=\cF(A)\cF(B)$;

3.~~~~~~~~ $\cF(I+rE_{ij})=1$ for $i\neq j$ and $r \in \quaternions$;\\
see \cite{ASL}. 
However, $~\mbox{Sdet}~$ {\em does not truly extend} $~\mbox{det}$.
Indeed, the two functionals do not coincide on complex matrices, 
since the former is nonnegative while the latter is truly complex.
In this paper we show that $\mbox{Sdet}$ {\em does extend the
nonnegative functional} $|$det$|$, namely the two functionals 
coincide for complex
matrices. More precisely, we show the following:

1) There exists no multiplicative functional on quaternionic 
matrices which coincides with $~\mbox{det}~$ on complex matrices.

2) $~\mbox{Sdet}~$ is the {\em only\ } non-constant multiplicative 
functional which coincides with $|$det$|$ on complex 
matrices (
we remark that just like $\mbox{det}[M]\neq 0$, the inequality 
$|\mbox{det}[M]|\neq 0$ characterizes non-singular matrices over
the complex numbers. The same central role in group theory over
the quaternions will be played by $\mbox{Sdet}[M]\neq 0$).

3) We show that the identities
$~|\mbox{det}(M)|=\prod |\lambda_i|~$, in terms of eigenvalues, and
$~|\mbox{det}(M)|=\prod \sigma_i~$, in terms of singular values,
extend to $M$ quaternionic. Thus, although $\mbox{Sdet}$ 
is originally defined through complexification~\cite{WIE}-\cite{TRA2}, 
it can be given concrete spectral and 
numerical-analytic interpretations which do not require
complexification.

4) We show that the Schur complements identity for complex matrices, 
$$|\mbox{det} \left[ \pmatrix{A&B\cr C&D} \right] |=
|\mbox{det} [A]|~|\mbox{det}[D-CA^{-1}B]|~,$$ 
extends to quaternionic matrices.

5) We discuss formulas
for $\mbox{det}[H],$ $\mbox{Adj}[H]$ and $H^{-1},$ based on the
classical permutation and minor calculation (some of this 
material can be found in \cite{CHEN}). It is interesting that
this approach, pursued by Cayley without success in the context
of {\em general} quaternionic matrices, is valid in the
{\em hermitian} case. The functional $\mbox{det}$, defined this 
way for hermitian quaternionic matrices, is {\em not} multiplicative.
Note that under the definition $\mbox{Sdet}[M]:=\mbox{det}[M^+M]$
one can extend the Study determinant to non-square matrices.

In the last section we also discuss some open 
problems concerning the behavior of the determinant 
and the difficulties of extending the formula 
$M^\inv=\mbox{Adj}(M)/\mbox{det}(M)$ to quaternions.



\section{Notation}

Quaternions, introduced by Hamilton~\cite{HAM1,HAM2} in 1843,  
can be represented
by four real quantities
\[
q=a+ i \, b + j \, c+ k \, d~,
~~~~~~a,b,c,d\in\reals~,
\]
and three imaginary units $i ,j,k$ satisfying
\[
i^2=j^2=k^2=ijk=-1~.
\]
We will denote by
\[\mbox{Re} [q] :=a~~~~~\mbox{and}~~~~~ 
\mbox{Im} [q] :=q-a =i \, b + j \, c+ k \, d~,\]
the real and imaginary parts of $q$.
The quaternion skew-field $\quaternions$ is an associative but 
non-commutative algebra of rank $4$ over $\reals$, endowed with
an involutory operation, called quaternionic conjugation,  
\[
\bar{q} = a-i \, b-j \, c-k \, d =
\mbox{Re} [q] - \mbox{Im} [q] ~,
\]
satisfying 
$\overline{pq}=\overline{q} \, \overline{p}$ for all 
$q,p \in \quaternions$. 
The quaternion norm $|q|$ is defined by
\[ 
|q|^2=q\bar q= a^2+b^2+c^2+d^2~.
\]
Among the properties of the norm, to be used in subsequent sections, we 
mention here the following
\[ |pq|=|qp|=|q| \, |p|~~~~~\mbox{and}~~~~~|1 - pq|=|1-qp|~.
\]
Every nonzero quaternion $q$ has a unique inverse 
\[
q^\inv=\bar q/|q|^2~.
\]
Two quaternions $p$ and $q$ are called similar if
\[
q=s^\inv p \, s~,~~~~~~~ \mbox{for some}~s\in\quaternions~.
\] 
By replacing $s$ by $u=s/|s|,$ we may always assume $s$ to be unitary. 
The usual complex conjugation
in $\complex$ may be obtained by choosing $s=j$ or $s=k$. A necessary 
and sufficient condition  for the similarity of 
$p$ and $q$ is given by
\[
\mbox{Re}[q] = \mbox{Re} [p]~~~~~\mbox{and}~~~~~
|\mbox{Im}[q]|=|\mbox{Im}[p]|~.
\] 
An equivalent condition is 
$Re[q]=Re[p]$ and $|q|=|p|.$
Every similarity class contains a complex number, unique up
to conjugation. Namely, every quaternion $q$ is similar to
$\mbox{Re}[q] \pm i \, |\mbox{Im}[q]|~.$
In particular, $q$ and $\bar{q}$
are similar. It can be seen that $s \in \quaternions$ conjugates
$q$ and $\bar q$ (i.e. $\bar q=s^\inv qs)$ if and only if $\mbox{Im}[q]=0$
or $ \mbox{Re} [qs] = \mbox{Re} [s] = 0$. 
However, there exists no fixed $s\in \quaternions$ which conjugates
$q$ and $\bar q$ for all $q\in\quaternions$.


\section{Spectral theory} 

Spectral theory for complex matrices admits several possible 
quaternionic extensions, which do not necessarily respect the 
fundamental theorem of algebra~\cite{EIN1,EIN2},
~\cite{NIV1}-\cite{EIL}. 
We shall be interested in the extension usually
described as ``right eigenvalues''~\cite{EIN1,EIN2},~\cite{BAK}.

Every $n \times n$ quaternion matrix $M$ is similar to 
an upper triangular matrix. This can be shown just as in 
the complex case. Using elementary Gaussian operations, the
general case can be reduced to the case of $2 \times 2$ matrices, 
where one wishes find $\alpha\in\quaternions$ so that 
\[
\pmatrix{\star & \star \cr 0 & \star } =
\pmatrix{1 & 0 \cr \alpha & 1 } 
\pmatrix{a & b \cr c & d } 
\pmatrix{1 & 0 \cr -\alpha & 1 }~, 
\]
given $a,b,c,d \in \quaternions$. By
permutation similarity we may assume that $b\neq 0.$ Solvability for
$\alpha$ is expressed by the non-commuting quadratic equation
\[
\alpha^{\2} b + \alpha (d-a) - c = 0~,
\]
which always has a solution~\cite{NIV1,NIV2}. 

Note that in the complex case the similarity matrix obtained 
in this procedure is not in general unitary; however, a different
procedure, Schur's lemma, triangularizes the matrix using unitary
similarity. Schur's lemma has been extended to quaternionic
matrices~\cite{KIP}. 

A modified version of the Jordan canonical form is valid
for quaternionic matrices. Namely, every matrix 
$M \in \quaternions^{\, \nxn}$ is similar, over the
quaternions, to a {\em complex} Jordan matrix $J$, 
defining a set of $n$ complex eigenvalues. However the 
eigenvalues $\lambda_i \in \complex$ are determined 
only up to complex conjugation~\cite{WIE}. 

The Schur and Jordan canonical forms are associated 
with {\em right eigenvalues} $M\psi = \psi \, q$, 
$\psi \in \quaternions^{\, \nx1}$, 
$q \in \quaternions$, which are determined only up to quaternionic 
similarity. This is further discussed 
in~\cite{EIN1,EIN2},~\cite{COH},~\cite{BAK}.

Let us denote by ${\cal Z}[M]$ the 
complexification~\cite{WIE},~\cite{TRA1,TRA2},~\cite{WOL} of the
quaternionic matrix $M$, i.e.
\begin{equation}\label{zm}
{\cal Z}[M] := \left( \begin{array}{cc}
M_{\1} & -M_{\2}^*\\ M_{\2} & M_{\1}^* \end{array} \right)~,
~~~~M = M_{\1} + j \, M_{\2}~,~~~M_{\1 ,\2} \in \complex^{\, \nxn}~.
\end{equation}
It has been shown in~\cite{WIE} that if $J$ is the complex Jordan 
form of $M$ then $J \oplus J^*$ is the Jordan form of ${\cal Z}[M]$.
Consequently the spectrum of ${\cal Z}[M]$ is 
$\left\{ \lambda_{\1}, \lambda_{\1}^*, \dots , \lambda_n, 
\lambda_n^* \right\}$.  


\section{On extending the determinant}
\label{s4}

The difficulty in extending the determinant to quaternions
results from lack of commutativity. Starting with Cayley himself
~\cite{CAY} all direct attempts at generalizing the concrete 
expression for the determinant have failed. Let us
consider the case of $2\times 2$ matrices.
Here, one may consider four different generalizations:
\begin{equation}
\label{wrong2}
ad-cb~,~~~~ad-bc~,~~~~da-cb~,~~~~da-bc~.
\end{equation}
It is easy to see that none of these expressions, alone or
jointly, has relevance to the invertibility of the matrix.
Consider, for example, the two matrices
\begin{equation}
\label{u1}
\mbox{$A~~=~~\frac{1}{\sqrt{2}}$} \, \left( \begin{array}{cc} 1 & i \\ j &k
\end{array} \right)~,~~~~~~~~~~
B~~=~~\mbox{$\frac{1}{\sqrt{2}}$} \,  \left( \begin{array}{cc} i & j \\ 
j & i 
\end{array} \right)~.
\end{equation}
In the case of $A$, exactly two expressions in~(\ref{wrong2}) vanish;
in the case of $B$, all the four expressions are zero. However,
both $A$ and $B$ are unitary.

In a different line of attack, one may look for multiplicative
functionals $\cF$ from $\quaternions^{\, \nxn}$ to $\quaternions$ which 
coincide with the determinant on complex matrices. Again, the result is 
negative:

{\it There is no multiplicative functional
\[ \cF \, : ~\quaternions^{\, \nxn}\rightarrow \quaternions~,\]
which coincides with   $\mbox{det}$
on complex diagonal matrices.} 

It is enough to obtain one counterexample for $n=2.$
Consider the $2 \times 2$ matrices
\[
M = \left( \begin{array}{cc} 1+i  & 0\\ 0 & i 
                                       \end{array} \right)~,~~~
N = \left( \begin{array}{cc} 1+i  & 0\\ 0 & -i 
                                       \end{array} \right)~,~~~
S = \left( \begin{array}{cc} 1  & 0\\ 0 & j
                                       \end{array} \right)~.
\]
Since $S$ is invertible, $\cF[S]\neq 0$, see Lemma \ref{LE1}. 
Since $SM=NS$, we conclude that  
$ \cF [S] \,  \cF [M] = \cF [N] \,  \cF [S]~,
$
hence $\cF [M]$ and $\cF [N]$ should be similar. This is a 
contradiction because obviously  
$\mbox{Re} \left\{ \cF [M] \right\} 
\neq \mbox{Re} \left\{ \cF [N] \right\}$.\\
$\Box$


\section{On extending the absolute value of the determinant}

A more positive result is obtained with respect to
the functional $|\mbox{det}|.$ 

\vspace*{.5cm}

\begin{theorem}
\label{t1} 
$Sdet$ is the unique functional 
\begin{equation}
\label{posi}
\cD \, : ~\quaternions^{\, \nxn}\rightarrow\reals_{\+}
\end{equation}
which is multiplicative, i.e.  
\begin{equation}
\label{mul}
\cD [MN] = \cD [NM]= \cD [M] \, 
\cD [N]~,
\end{equation}
and satisfies the scaling condition
\begin{equation}
\label{qI}
\cD [q I] = |q|^n~,~~~\forall q \in \quaternions~.
\end{equation}
\end{theorem}

\vspace*{.5cm}

Before proving this theorem, we make some observations
concerning nonnegative multiplicative functionals. The only
non-trivial part here is the last property, which has been
proved elsewhere.

\begin{lemma}
\label{LE1}
If $\cal{F} ~:~ \quaternions^{\, \nxn}\rightarrow\reals_{\+}$ 
is a non-constant multiplicative functional, then:\\

1) $ \cF[S]  = 1$, if $S^{\2}=I$;\\

2) $\cF[S]  \cF[S^\inv] = 1$ and $\cF[S^\inv M S]  = \cF[M]$, if $S$ is invertible;\\

3) $\cF[P] = 1$ for all permutation matrix  $P$;\\

4) $\cF[M] = 0$ if and only if $M$ is singular.
\end{lemma}

\vspace*{.5cm}

\noindent
{\em Proof.}\\

Multiplicativity and non-triviality implies that $\cF[I]=1.$
Now items 1-2 become trivial consequences of multiplicativity. 
Item 3 follows from the fact that every permutation matrix is 
a product of elementary permutation matrices $P_i$ with $P_i^2=I.$
As for item 4, if $M$ is not singular then applying $\cF$ to
$MM^\inv=I$ implies that $\cF[M]\neq 0.$ If $M$ is singular, 
$\cF\neq 0$ by a result of~\cite{BRE}, see~\cite{ASL} pag.~58.\\
$\Box$

\vspace*{.5cm}

\noindent
{\em Proof of Theorem \ref{t1}}.\\

Let $\cD$ be a functional satisfying (\ref{posi},\ref{mul}, \ref{qI}).
Let $\{ E_{ij} \}_{i,j=1}^n$ be the usual canonical basis 
over $\quaternions$ in $\quaternions^{\, \nxn}$. Let $q\in\quaternions$
be non-zero.  Consider the $n$ diagonal elementary matrices $M_i(q)$
\[
M_i(q) := I + (q-1) \, E_{ii}~
\]
and the $n(n-1)/2$ upper triangular elementary matrices
\[
M_{ij}(q) := I + q \, E_{ij}~,~~~i < j~.
\]
First we show that
\begin{equation}\label{diag}
\cF [ M_i (q)] = |q|~.
\end{equation}
Indeed, by permutation similarity we see that $\cF [ M_i(q) ]$ is
independent of $1 \leq i \leq n$. So set $f(q):=\cF [ M_i(q) ]$.
We have $q I = \prod_{i=1}^{n} M_i (q)$, hence
\[
|q|^n = \cF [qI] = \prod_{i=1}^{n} \cF [ M_i (q)] = f^n(q)~.
\]
Hence $f(q) = |q|$, as claimed. Next we show that
\begin{equation}\label{tria}
\cF [ M_{ij}(q)] = 1~.
\end{equation}
Indeed, it is easy to see that
$$M^{\inv}_{ij}(q) = M_{ij}(-q)=M_i(-1) M_{ij}(q) M_i (-1),$$ 
hence $\cF [ M^{\inv}_{ij}(q)] = \cF [ M_{ij}(q)]$. Now (\ref{tria})
follows by multiplicativity.

We have therefore established that
$\cF : \quaternions^{\, \nxn}\rightarrow\quaternions$
satisfies the three Dieudonn\'e conditions (\ref{mul}), (\ref{tria})
and item {\em 4} in Lemma \ref{LE1}. 
Therefore, according to Dieudonn\'e's result~\cite{ASL}  
$\cF=\mbox{Ddet}^{2r}=\mbox{Sdet}^{r}$ for some $r\in\reals$.
Due to (\ref{qI}), it is readily seen that $r=1.$\\
$\Box$

Note that in general if $~\cF~$ is multiplicative and 
$~r\in\reals~$ then~ $\cF^r~$ is also multiplicative. 
Therefore, we have a one-parameter group of
nonnegative multiplicative functionals: $\{\mbox{Sdet}^r:~r\in\reals\}$
(The case $r=0$ is interesting: it leads to the functional
whose value is $1$ on all the invertible matrices and 
$0$ otherwise). In view of Theorem (\ref{t1}) we conclude 
that this one-parameter family, plus the 
two constant functionals $\cF_0[M] \equiv 0$ and $\cF_1[M] \equiv 1,$ 
are the only nonnegative multiplicative functionals on
quaternionic matrices.

\section{Concrete description of the Study determinant} 

Theorem \ref{t1} has the following main corollaries:

\begin{corollary} \label{111} {\em
If $M$ is upper triangular then $\mbox{Sdet}(M)=\prod_{i=1}^n 
\left| M_{ii} \right|.$} 
\end{corollary}

This follows easily by writing $M$ explicitly as a product of
elementary matrices, using
(\ref{diag},\ref{tria}).\\
$\Box$

\begin{corollary}
\label{evs}
{\em 
For all matrix $M$, $\mbox{Sdet}(M) = \prod_{i=1}^{n} | \lambda_i |~$
where $\lambda_i$ are the eigenvalues of $M$.}
\end{corollary}

By Lemma \ref{LE1} item 2, it is enough to consider the Jordan form,
or the Schur form, of $M$ which is of the type considered by
Corollary \ref{111}.\\
$\Box$

Since the eigenvalue identity just exhibited, restricted to 
complex matrices, is also valid for $|\mbox{det}|,$ we get immediately:

\begin{corollary} {\em
For complex matrices we have $\mbox{Sdet}(M) = |\mbox{det}(M)|~.$}
\end{corollary}

Let us define the adjoint of $M$ by 
$\left( M^{\+} \right)_{ij} =  \overline{M_{ji}}$.
A matrix $U \in \quaternions^{\, \nxn}$ is called unitary if 
$U^{\+}U=I$. According to the quaternionic Schur lemma~\cite{KIP}, 
every $n \times n$ quaternionic matrix, $M$, can be written as 
$M=U^{\+} T U$ where $U$ is  unitary and $T$ triangular. 
Since in addition Eq.~(\ref{tria}) is obviously
valid for lower as well as upper triangular matrices, we get:

\vspace*{.5cm}

\begin{corollary}
{\em 
$\mbox{Sdet}(M^{\+}) = \mbox{Sdet}(M)$. In particular 
$\mbox{Sdet}(U)=1$ if $U$ is unitary.}
\end{corollary}

The identity $\mbox{Sdet}(M)=1$ may be taken as a basis to define
the group of {\it unimodular} matrices.

\vspace*{.5cm}

Next, we calculate $\cF$ in terms of singular values. 
The singular value decomposition, SVD, 
for complex matrices extends to quaternionic matrices in a 
straightforward way. Every $n \times n$
quaternionic matrix, $M$, has the SVD $M=U\Sigma V$ where 
$U$ and  $V$  are unitary, $\Sigma = \Sigma_{\1} \oplus 0$,  
$\Sigma_{\1} = \sigma_{\1} \oplus \dots \oplus \sigma_{k}$, where
$\sigma_{\1} \geq \sigma_{\2} \geq \dots \sigma_k \geq 0$ are the singular
values of $M$~\cite{WIE,KIP,LEE,BRE2}. In these terms the following holds:

\vspace*{.5cm}

\begin{corollary}
\label{sig} 
$\cF [M]= \prod_{i=\1}^{n}  \sigma_{i}$.
\end{corollary}

\section{ Hermitian matrices}

A quaternionic matrix $H$ is called hermitian if 
$H^{\+}= H$. As we saw in section~(\ref{s4}),  
the common determinant cannot be extended to
quaternionic matrices. However, it can be extended to {\em hermitian}
quaternionic matrices. The usual definition of the determinant in terms
of permutations was generalized in the Chinese literature, see for
example ref.~\cite{CHEN}. Another possible definition is analogous
to Corollaries \ref{evs} and \ref{sig}:
\[ \left| H \right|_r = \prod_{i=1}^{n}  \lambda_{i}~.\] 
Note that for hermitian matrices   
the eigenvalues are uniquely determined and real. This follows
from the fact that ${\cal Z}[M]$ is also hermitian.  
Note that 
the set of hermitian matrices is not closed 
under products, and the functional $det:H\rightarrow|H|_r$ is 
not multiplicative. However, it is invariant under congruence.

It is easy to show that for hermitian matrices 
the following are equivalent:\\
1) $H$ is positive definite, i.e. $x^{\+} H x > 0$ for all non zero 
$x \in \quaternions^{\, \nx1}$;\\
2. All the eigenvalues $\lambda_i$ are positive;\\
3. All the (signed) {\em real determinants} of the 
principal minors are positive. 
 
We conclude this section by comparing  the functional
$\mbox{Sdet}[M]$, the functional $\left| H \right|_r$ just defined,
and the {\em q-determinant}~\cite{ZHA}  
\[
\left| M \right|_q = \mbox{det} \left\{ {\cal Z} \left[ M \right] 
\right\}~,
\]
when ${\cal Z} \left[ M \right]$ is defined in equation (\ref{zm}). 
From previous considerations, we have
\[
\left| M \right|_q = 
 \prod_{i=\1}^{n}  |\lambda_{i}|^{\2} = 
\mbox{Sdet}[M^{\+}] \, \mbox{Sdet}[M] =\mbox{Sdet}^{\2}[M] = 
\left| M^{\+} M \right|_r~. 
\]
Using this equation, one can extend the definition of Sdet from
square to non square matrices. This approach is found in~\cite{CHEN},
where the resulting functional is called {\em double determinant}.


\section{ Schur complements}

Let ${\cal R}$ be an associative ring.  A matrix $M\in {\cal R}^{\, \nxn}$ is 
called invertible if $MN=NM=I_{n}$ for some $N\in {\cal R}^{\, \nxn}$, which
is necessarily unique. It is shown in~\cite{ZHA} that
in case ${\cal R} = \quaternions$, 
$MN=I_{n}$ implies $NM=I_{n}$. 

The Schur complements procedure~\cite{HJ} is a powerful  
tool in calculating inverses of matrices over rings. 
Let us write a generic $n$-dimensional matrix $M$
in block form 
\[
M=\left( \begin{array}{cc}A&B\\ C&D\end{array} \right)~.
\]
Assuming that $A\in {\cal R}^{\kxk}$ is invertible, one has 
\begin{equation}\label{M=}
M=
\left( \begin{array}{cc} I_{k} & 0 \\ C A^\inv & I_{n-k} \end{array} \right)
\left( \begin{array}{cc} A & 0 \\ 0 & A_{\S} \end{array} \right) 
\left( \begin{array}{cc}I_{k} &A^\inv B\\ 0&I_{n-k}
\end{array} \right)~,
\end{equation}
with
\[
A_{\S} := D-CA^\inv B ~.
\]
We shall call $A_{\S}$ the {\em Schur complement} of $A$ in $M$.

The invertibility of $A$ ensures 
that the matrix $M$ is invertible if and 
only if $A_{\S}$ is invertible, and the inverse is given by 
\begin{equation}
\label{d}
M^\inv=\left( \begin{array}{cc}I_{k}&-A^\inv B\\ 0&I_{n-k}\end{array} \right)
\left( \begin{array}{cc}A^\inv&0\\ 0&{A_{\S}}^\inv \end{array} \right)
\left( \begin{array}{cc}I_{k}&0\\ -C A^\inv&I_{n-k}\end{array} \right)~.
\end{equation}
The inversion of an $n$-dimensional matrix is thus reduced to inversion
of two smaller matrices,
$A \in {\cal R}^{\kxk}$ and $A_{\S} \in {\cal R}^{\, \nkxnk}$
(plus some multiplications); repeated use of this size 
reduction can be used to invert the matrix efficiently.
It is not as efficient as Gaussian elimination, but the
latter may not be available in general rings.


\vspace*{.5cm}

\begin{corollary}
\label{71}
{\em  
$\mbox{Sdet} \left[ \left( \begin{array}{cc} A & B\\ C & D 
                                       \end{array} \right) \right] =
\mbox{Sdet}[A] \, \mbox{Sdet} [D - C A^\inv B]$ ~as long as
$A^\inv$ exists.}
\end{corollary}

Indeed, from the construction of $\mbox{Sdet}$ in the last section,
we see that its value on each of the two block-triangular 
matrices in (\ref{M=}) is $1$; since the eigenvalues of a
direct sum are the union of the eigenvalues of the summands,
we get that $\mbox{Sdet}[A\oplus A_S]=\mbox{Sdet}[A]\mbox{Sdet}[A_S].$
This plus multiplicativity implies the result.\\
$\Box$
 
\vspace*{.5cm}

As a result of the Schur complements determinant formula just
exhibited, we get the following commutation formula for $\mbox{Sdet}$,
which generalizes a well known property of $\mbox{det}$ (actually,
of $|\mbox{det}|)$:

\begin{corollary} {\em 
$\mbox{Sdet} [ I + M N] = \mbox{Sdet} [ I + N M]~$
for all $~M \in \quaternions^{\nxm}~$ and 
$~N \in \quaternions^{\mxn}$.}
\end{corollary}

Indeed, consider the matrix
\[ 
\left( \begin{array}{cc} I_{\1} & N \\ M  & I_{\2} 
\end{array} \right) 
\] 
and apply to it Schur complements with respect to both $I_{\1}$ and
$I_{\2}$, respectively. We get
\[
\mbox{Sdet} \left[ \left( \begin{array}{cc} I_{\1} & N \\ M  & I_{\2} 
\end{array} \right) \right]
= \mbox{Sdet} [I_{\1}] \, 
\mbox{Sdet}  [I_{\2} - M I_{\1}^\inv N]~,
\]
and
\[
\mbox{Sdet} \left[ \left( \begin{array}{cc} I_{\1} & N \\ M  & I_{\2} 
\end{array} \right) \right] = \mbox{Sdet} [I_{\2}] \, 
\mbox{Sdet} [I_{\2} - N I_{\1}^\inv M]~,
\]
implying the identity.\\
$\Box$
 
\vspace*{.5cm}   


\section{The case of $2\times 2$ matrices} 

In this last section, we discuss inversion, adjoint and 
determinant  for $2 \times 2$ quaternionic matrices. 

\subsection*{$\bullet$ Inversion} Let
\[ M = \left( \begin{array}{cc}a&b\\ c&d\end{array} \right) 
\] 
be an invertible $2\times 2$ matrix with quaternionic entries. 
When $a$, $b$, $c$, $d$  are all non-zero, four parallel 
applications of the Schur complement formula (\ref{d}) lead 
to a concrete description of the inverse:
\begin{equation}
\label{inv.2x2}
M^\inv=\left( \begin{array}{cc}\tilde a&\tilde b\\ \tilde c&\tilde d\end{array} \right)~,
\end{equation}
where
\begin{equation}
\label{e4}
\begin{array}{ll}
\tilde a=(a-bd^\inv c)^\inv,&~~~\tilde b=(c-db^\inv a)^\inv~,\\
\tilde c=(b-ac^\inv d)^\inv~,&~~~\tilde d=(d-ca^\inv b)^\inv~,
\end{array}
\end{equation}
see G\"ursey~\cite{GUR} page 115.
The invertibility of $M$ guarantees that these four values 
are well-defined non-zero quaternions.
What happens if some of the entries of $M$ vanish? 
Assume for example that $a=0$. 
The invertibility of $M$ implies that $b,c\neq 0$. Consequently, 
the element $d-ca^\inv b$ has {\em infinite} modulus. In this case, we
define
\[ 
\tilde d:=\lim_{a \rightarrow 0}\, 
(d-ca^\inv b)^\inv~.
\]
A simple  calculation, 
$$| \tilde d |  := 
\lim_{a \rightarrow 0} \, 
\frac{1}{|d-ca^\inv b|} =  
\lim_{a \rightarrow 0} \, 
\frac{1}{|c| \, |c^\inv d- a^\inv b|} =$$
$$ = \lim_{a \rightarrow 0} \, 
\frac{|a|}{|c| \, |a c^\inv d - b|}
 =  \lim_{a \rightarrow 0} \, 
\frac{|a|}{|c| \, |b|}  =  0~,$$
shows that $\tilde{d} = 0$. Thus,
\[
\left( \begin{array}{cc}0 & b \\ c & d\end{array} \right)^\inv
=
\left( \begin{array}{cc}\tilde a&\tilde b\\ \tilde c& 0 \end{array} \right)~,~~~
\tilde{a}= - c^\inv  d  b^\inv \, ,~\tilde{b} = c^\inv \, ,~
\tilde{c} = b^\inv~.
\]
We conclude that Eq.~(\ref{e4}) remain valid under appropriate 
conventions, when some entries in $M$ are zero. We do not have a clear 
generalization of this phenomenon for $n>2$.

\subsection*{$\bullet$ Adjoint}
Eqs.\,(\ref{inv.2x2}, \ref{e4}) are valid in every associative 
ring ${\cal R}$. In
case ${\cal R}$ is also commutative,  
Eqs.\,(\ref{inv.2x2},\ref{e4})
reduce to the well known formula 
\begin{equation}
\label{adj}
M^\inv=\frac{\mbox{Adj}[M]}{\mbox{det} [M]}~.
\end{equation}
In calculating the inverse of real and complex matrices, (\ref{adj})
is of great theoretical importance. So far, we have failed to 
generalize this formula to quaternion matrices. At first sight, it might
make sense to conjecture a non-commuting expression of the general form,
\begin{equation}
\label{gen}
M^\inv = P \,  \mbox{Adj} [M]  \, Q~,
\end{equation}
with quaternionic diagonal matrices
$P=\mbox{diag} \left\{ p_{\1} , p_{\2} \right\}$ and 
$Q=\mbox{diag} \left\{ q_{\1} , q_{\2} \right\}$.
Nevertheless, the resulting constraints
\[
\begin{array}{ccrcccr}
p_{\1} & = &   \tilde{a} q_{\1} d^{\inv}~, & ~~~ &
p_{\2} & = & - \tilde{c} q_{\1} c^{\inv}~,\\
p_{\1} & = & - \tilde{b} q_{\2} b^{\inv}~, & ~~~ &
p_{\2} & = &   \tilde{d} q_{\1} a^{\inv}~,
\end{array}
\]
which, for commutative fields, are satisfied if 
$P =\mbox{det}^{-1} [M]~I$ and $Q=I,$  
are not always solvable. For example, the first matrix in (\ref{u1}) 
cannot be written in the form~(\ref{gen}). Whether a further weakening, 
beyond~(\ref{gen}), of formula ~(\ref{adj}) is valid for quaternionic 
matrices remains an open problem. The mere definition of $\mbox{Adj}[M]$,
$M \in \quaternions^{\, \nxn}$, $n>2$, preserving~(\ref{adj}),  
is not clear.

A different generalization of~(\ref{adj}) for $2 \times 2$ quaternionic 
matrices may be obtained using a Hadamard product between a 
non negative matrix and a termwise-unitary quaternionic matrix 
\begin{equation}
M^{-1} = \frac{1}{\mbox{Sdet}[M]} \, \left( \begin{array}{cc}
|d| & |b|\\
|c| & |a| \end{array} \right) \, \circ \,
\left( \begin{array}{cc}
\frac{\tilde{a}}{|\tilde{a}|}  &  \frac{\tilde{b}}{|\tilde{b}|}\\
\frac{\tilde{c}}{|\tilde{c}|}  & \frac{\tilde{d}}{|\tilde{d}|} 
\end{array} \right)~.
\end{equation}

Another description of the inverse matrix is offered in equation (37)
of the paper of Chen~\cite{CHEN}.

\subsection*{$\bullet$ Determinant} For $n=2$, it is noteworthy that
the following four quaternion expressions are equal:

\[
|a| \, |d - c a^\inv b | = 
|b| \, |c - d b^\inv a | = 
|c| \, |b - a c^\inv d | = 
|d| \, |a - b d^\inv c |~.
\]
From the Schur complement formula, Corollary \ref{71}, it follows that
each of this expressions, properly extended in case $a,b,c$ or $d$
is zero, expresses the value of $\mbox{Sdet}\pmatrix{a&b\cr c&d}.$
Applying this formula on the two unitary matrices in (\ref{u1}),
one obtains the expected result (these matrices are
unitary, hence unimodular).
For hermitian quaternionic matrices, the real 
determinant is given by
\[
\left| \left( \begin{array}{cc} \alpha & q\\ \bar{q} & \delta 
                                       \end{array} \right) \right|_r =
\lambda_{\1} \, \lambda_{\2} = \alpha \, \delta - |q|^{\2}~,~~~~~
\alpha \, , \, \delta  \, \in \reals~,~~~q \in \quaternions~. 
\]

\end{document}